\documentclass[epjCONF]{svjour}
\usepackage{graphics}
\usepackage[varg]{txfonts} 
\usepackage[latin1]{inputenc}
\session-title{Conference Title, to be filled}
\begin{document}
\title{Heavy and light quarks in the instanton vacuum }
\author{M. Musakhanov\thanks{\email{yousuf@uzsci.net}}}
\institute{National University of Uzbekistan}
\abstract{
Assuming the gluon field is well approximated by instanton configurations we derive a light quarks determinant and calculate its contribution to the specific heavy quarks correlators -- namely, the  heavy quark propagator and heavy quark-aniquark correlator, receiving the instanton generated light-heavy quarks interaction terms contributions.
With these knowledge we calculate the light quark contribution to the interaction between heavy quarks, which might be essential for the properties of a few heavy quarks systems.
}

\maketitle
\section{Introduction.}
\label{intro}
The physics of the heavy mesons and baryons with open and hidden heavy
quarks is very reach and hot topic. Understanding the heavy-meson
physics is important for evaluation of the components of the $CKM$-matrix,
verification of the Standard Model and probing the physics beyond
it, as well as production of different exotic meson states. Currently
the experiments with $B$- and $D$-mesons are intensively studied
by Belle~\cite{Belle}, 
BaBar~\cite{BABAR}
and CDF collaborations, where unprecedented integrated luminocities
were achieved, as well as neutrino-production of open and hidden charm
in neutrino-hadron processes studied by K2K~\cite{K2K},
MiniBoone~\cite{MiniBooNE}, 
NuTeV~\cite{NuTeV}
and Minerva~\cite{Minerva} collaborations.

Theoretically, in pre-QCD era some success was achieved by the quantum-mechanical
models which use effective potentials to describe heavy hadrons and
their excitations (see e.g.~\cite{Eichten:1979ms} and references
therein). However, such description inevitably introduces undefined
phenomenological constants. The relation of these constants to QCD
parameters is quite obscure: due to interaction with gluons and virtual
light quark pairs all the constants contain nonperturbative dynamics.
The numerical values of these constants are determined from fits to
experimental data, which limits the predictive power of such models. 

An advanced version of the potential model is NRQCD \cite{Bodwin:1994jh},
however in this model light quarks and their interactions with heavy
quarks via gluons is done in a phenomenological way. For this reason
it is limited to description of systems with two heavy quarks. Alternatively,
the heavy mesons are described in the Heavy Quark Effective Theory
(HQET) proposed in~\cite{Isgur:1989}, which treats
the heavy mesons using the pQCD methods but does not take into account
nonperturbative effects.

We propose to study the heavy quark physics in
the framework of the instanton vacuum model. This model was developed
in~\cite{Diakonov} and provided
a consistent description of the light mesons physics~\cite{Musakhanov}.

One of the most prominent advances of the instanton vacuum model 
is the correct description of the spontaneous breaking of the chiral
symmetry ($S\chi$SB), which is responsible for properties of most
hadrons and nuclei ~\cite{Leutwyler:2001hn}. The $S\chi$SB is due
to specific properties of QCD vacuum, which  is known to be one of
the most complicated objects due to perturbative as well as non-perturbative
fluctuations and is a very important object of investigations by methods
of Nonperturbative Quantum Chromo Dynamics (NQCD).  In the instanton
picture  $S\chi$SB is due to the delocalization of single-instanton
quark zero modes in the instanton medium. One of the advantages of
the  instanton vacuum is that it is characterized by only two parameters:
the average instanton size $\rho\sim0.3\,{\rm fm}$ and the average
inter-instanton distance $R\sim1\,{\rm fm}$. These essential numbers
were suggested in~\cite{Shuryak:1981ff} and were derived from $\Lambda_{\overline{{\rm MS}}}$
in ~\cite{Diakonov}. These values were recently confirmed
by lattice measurements \cite{lattice}.

In case of the heavy quarks, the instanton vacuum description was
discussed in~\cite{Diakonov:1989un,Chernyshev:1995gj}. For the heavy quarks 
even the charmed quark mass $m_{c}\sim1.5$~GeV is larger than the typical parameters
of the instanton media--the inverse instanton size $\rho^{-1}\approx600$~MeV
and the interinstanton distance $R^{-1}\approx200$~MeV and thus the quark mass determines the 
dynamics of the heavy quarks.

\section{Light quark determinant with the quark sources term.}
\label{Light quark}

 Instanton vacuum field is assumed as a superposition of $N_{+}$
instantons and $N_{-}$ antiinstantons: \begin{eqnarray}
A_{\mu}(x)=\sum_{I}^{N_{+}}A_{\mu}^{I}(\xi_{I},x)+\sum_{A}^{N_{-}}A_{\mu}^{A}(\xi_{A},x).\label{A}\end{eqnarray}
 Here $\xi=(\rho,z,U)$ are (anti)instanton collective coordinates--
size, position and color orientation (see reviews ~\cite{Diakonov,Schafer:1996wv}.
The main parameters of the model are the average inter-instanton distance
$R$ and the average instanton size $\rho$. The estimates of these
quantities are \begin{eqnarray}
 &  & \rho\simeq0.33\, fm,\, R\simeq1\,{\rm fm},\mbox{(phenomenological)}~~\mbox{\cite{Diakonov,Schafer:1996wv}},\nonumber \\
 &  & \rho\simeq0.35\, fm,\, R\simeq0.95\,{\rm fm},\mbox{(variational)}~~\mbox{\cite{Diakonov}},\nonumber \\
 &  & \rho\simeq0.36\, fm,\, R\simeq0.89\,{\rm fm},~\mbox{(lattice)}~\mbox{\cite{lattice}}\label{classicalParameters}\end{eqnarray}
 and have $\sim10-15\%$ uncertainty.

Our main approximation is the interpolation formula for the light quark propagator in a single instanton field: 
\begin{eqnarray}\label{Si}
&&S_{i}=S_{0}+S_{0}\hat{p}\frac{|\Phi_{0i}><\Phi_{0i}|}{c_{i}}\hat{p}S_{0},\\ \nonumber
&&S_{0}=\frac{1}{\hat{p}+im},\,\,\,
c_{i}=im<\Phi_{0i}|\hat{p}S_{0}|\Phi_{0i}>\,.
\end{eqnarray}
 The advantage of this interpolation is shown by the projection of
$S_{i}$ to the zero-modes: \begin{eqnarray}
S_{i}|\Phi_{0i}>=\frac{1}{im}|\Phi_{0i}>,\,\,\,<\Phi_{0i}|S_{i}=<\Phi_{0i}|\frac{1}{im}\end{eqnarray}
 as it must be, while the similar projection of $S_{i}$ given by
~\cite{Diakonov} has a wrong component, negligible only in
the $m\rightarrow0$ limit.

Summation of the re-scattering series leads to the light quark propagator in the instanton vacuum:
\begin{equation}
S =S_{0} -S_{0}\sum_{i,j}\hat p |\Phi_{0i}>
<\Phi_{0i}|\frac{1}{B}|\Phi_{0j}>
<\Phi_{0j}|\hat p  S_{0},
\label{propagator1}
\end{equation}
where $B=\hat pS_0\hat p.$
Here $\tilde {\rm Tr}$ means the trace on the flavor and only on 
zero-mode ($|\Phi_{0j}> $) space.
The  explicit form of the matrix $ B(m)$ on the flavor and only on 
zero-modes ($|\Phi_{0j}> $) space is:
\begin{equation}
B^{fg}_{ij}=\delta_{fg}
<\Phi_{0i}|\hat p  \, S_{0,f}
\hat p |\Phi_{0j}> .
\end{equation}
Then the low-frequency part of the light quark determinant~\cite{Musakhanov} is
\begin{equation}
 {{\rm Det}}_{\rm low} [m] = {\rm det} B(m).
\label{detB}
\end{equation}
Making few further steps~\cite{Musakhanov} we get 
the fermionized representation of low-frequencies
light quark determinant in the presence of the quark sources, which is relevant for our problems, in the
form:  
\begin{eqnarray}
&&{\rm Det}_{\rm low}\exp(-\xi^{+}S\xi)=\int\prod_{f}D\psi_{f}D\psi_{f}^{\dagger}  \prod_{\pm,f}^{N_{\pm}}V_{\pm,f}[\psi^{\dagger},\psi]
 \nonumber\\\label{part-func}
&&\times\exp\int\sum_{f}\left(\psi_{f}^{\dagger}(\hat{p}\,+\, im_{f})\psi_{f}+\psi_{f}^{\dagger}\xi_{f}+\xi_{f}^{+}\psi_{f}\right),
\end{eqnarray}
 where 
 \begin{eqnarray}\label{V}
V_{\pm,f}[\psi^{\dagger},\psi]=&&i\int d^{4}x\left(\psi_{f}^{\dagger}(x)\,\hat{p}\Phi_{\pm,0}(x;\zeta_{\pm})\right)\\\nonumber
&&\times\int d^{4}y\left(\Phi_{\pm,0}^{\dagger}(y;\zeta_{\pm})(\hat{p}\,\psi_{f}(y)\right).
\end{eqnarray}
   The light quark partition function $Z[\xi_f,\xi_f^+]$ is given by the averaging of ${\rm Det}_{\rm low}\exp(-\xi^{+}S\xi)$ over the collective coordinates of the instantons $\zeta_\pm$ as:
\begin{eqnarray}\nonumber
Z[\xi_f,\xi_f^+]=\int D\zeta &{\rm Det}_{\rm low}\exp(-\xi^{+}S\xi),\,\,\, D\zeta=\prod_\pm d\zeta_\pm.
\end{eqnarray}
The averaging over collective coordinates $\zeta_{\pm}$ is a rather simple procedure,
since factorized form of the Eq.~(\ref{part-func})
and the low density of the instantons ($\pi^{2}\left(\frac{\rho}{R}\right)^{4}\sim0.1$).
These one allows us to average over positions and orientations of the instantons independently. 

Light quark partition function at $N_f=1$ and $N_\pm=N/2$ is exactly given by 
\begin{eqnarray}
&&Z[\xi,\xi^+]=\exp\left[{-\xi^+\left(\hat p \,+\, i(m+M(p))\right)^{-1}\xi}\right]\label{Z}\\\nonumber
&&\times\exp\left[{\rm Tr}\ln\frac{\hat p+im+iM(p)}{\hat p+im }+N\ln\frac{N/2}{\lambda}-N\right]
\\
&&N={\rm tr}\frac{iM(p)}{\hat p \,+\, i(m+M(p))},\, M(p)=\frac{\lambda}{N_c}(2\pi\rho F(p))^2.
\label{M}
\end{eqnarray}
Here the form-factor $F(p)$ is given by Fourier-transform of the zero-mode. The coupling  $\lambda$ and the dynamical quark mass $M(p)$ are  defined by the Eq. (\ref{M}).

At $N_f=2$, $N_\pm=N/2$ and saddle-point approximation (no meson loops contribution)
\begin{eqnarray}
&&Z[\xi_f,\xi_f^+]=\exp\left[-\sum_f\xi_f^+\left(\hat p+im_f+iM_f(p)\right)^{-1}\xi_f\right]
\label{ZNf=2}\\\nonumber
&&\times\exp\left[N\ln\frac{N/2}{\lambda}-N
- \frac{V\sigma^2}{2}+ \sum_f{\rm Tr}\ln\frac{\hat p+im_f+iM_f(p)}{\hat p+im_f }\right].
\end{eqnarray}
Here $\lambda,\sigma$ and dynamical quark mass 
$$M(p)=\frac{\lambda^{0.5}}{2g}( 2\pi\rho)^2F^2(p)\sigma,\,\,\,g^2=\frac{(N^2_c-1)2N_c}{2N_c-1}$$  
are defined from the Eqs.
\begin{eqnarray}
N=\frac{1}{2}{\rm Tr}\frac{iM_f(p)}{\hat p+im_f+iM_f(p)}=\frac{1}{2}\sigma^2.
\end{eqnarray}
In general, at $N_f >2$,  and in the saddle-point approximation (no meson loops contribution) $Z[\xi_f,\xi_f^+]$ has a similar form as the Eqs. (\ref{Z}, \ref{ZNf=2}). 

 \section{Light quark propagator }
\label{Light quark propagator}

The propagator is defined as
\begin{eqnarray}
S=\int DA \,\,{\rm Det}\left(\hat P+im\right)\frac{1}{\hat P+im},\,\,\, \hat P=\hat p+\hat A.
\end{eqnarray}
In the instanton vacuum model $A\approx\sum_i A_i$, where $A_i$ are instantons and $DA\approx D\zeta .$  
Then, accordingly Eq.~(\ref{M}) the light quark propagator is:
\begin{eqnarray}
S=\frac{1}{\hat p \,+\, i(m+M(p))}.
\end{eqnarray}
Pobylitsa~\cite{Pobylitsa:1989uq} neglected by the quark determinant:
\begin{eqnarray}
S_{Pob}=\int D\zeta \frac{1}{\hat P+im}  
\end{eqnarray}
and derived the Eq.:
\begin{eqnarray}
S^{-1}_{Pob}= S_0^{-1} +\int D\zeta  \sum_i(S_{Pob}-\hat A^{-1}_i)^{-1},
\label{PobEq}
\end{eqnarray}
where it was applied large $N_c$ argumentation. Representing $S^{-1}_{Pob}- S_0^{-1} = \Sigma$,
it was found the Eq.:
\begin{eqnarray}
&&\Sigma
=\frac{N}{2VN_c}{\rm tr_c} \sum_\pm \int dz_\pm \frac{ \hat p|\Phi_{0,\pm}> <\Phi_{0,\pm}| \hat p}{\Sigma_0}+O\left[\left(\frac{N}{VN_c}\right)^2\right], 
\nonumber\\
&&\Sigma_0=<\Phi_{0,\pm}|\Sigma|\Phi_{0,\pm}>
\end{eqnarray}
and
finally ($m=0$ case) the solution for the dynamical quark mass:
\begin{eqnarray}\label{MPob}
M^2_{Pob}(k)= \frac{N}{4VN_c}\frac{(2\pi\rho)^4 F^4(k)}{ \int\frac{d^4q}{(2\pi)^4}\frac{(2\pi\rho)^4 F^4(q)}{q^2}}
\end{eqnarray}
corresponding to the Eq.
\begin{eqnarray}\label{MPob1}
4N_c \int\frac{d^4q}{(2\pi)^4}\frac{M^2_{Pob}(q)}{q^2}=\frac{N}{V}
\end{eqnarray}
The Eq. (\ref{M}) for the $\lambda$ ($m=0$ case) has explicit form
\begin{eqnarray}\label{Mdet}
4N_c \int\frac{d^4q}{(2\pi)^4}\frac{M^2(q)}{q^2+M^2(q)}=\frac{N}{V}.
\end{eqnarray}
As we see,  the difference between Eqs. (\ref{MPob1}) and  (\ref{Mdet}) is only in the denominators.
This one is due to the account of the quark determinant in the derivation of the Eq.~(\ref{Mdet}) (and Eq.~(\ref{M})). 
In the following we will use the Eq. (\ref{M}) for the dynamical quark mass $M(k)$.

Any $N_f$ case in the saddle-point approximation has no essential difference with the present case $N_f=1$.

\section{Heavy quark propagator.}
\label{Heavy quark}
At the ref.~\cite{Diakonov:1989un}  it was considered the Eq. for the heavy quark propagator in the line similar the Eq.~(\ref{PobEq}).
Our aim here is to extend the approach~\cite{Diakonov:1989un} taking in-to account the light quarks contribution at $N_f=1$ case.
So, define the heavy quark propagator as: 
\begin{eqnarray}
&&S_H=\frac{1}{Z}\int D\psi D\psi^{\dagger} 
\prod_{\pm}^{N_{\pm}}\bar V_{\pm}[\psi^{\dagger} ,\psi ]\,e^{\int\psi^{\dagger}(\hat p+im )\psi} w[\psi,\psi^\dagger],
\nonumber\\ \nonumber
&&w[\psi,\psi^\dagger]
=\int \frac{D\zeta}{\prod_{\pm}^{N_{\pm}}\bar V_{\pm}[\psi^{\dagger} ,\psi ]}
\prod_{\pm}^{N_{\pm}}V_{\pm}[\psi^{\dagger} ,\psi ]\frac{1}{\theta^{-1}-\sum_i a_i},
\nonumber\\
&& w_\pm=\frac{1}{\theta^{-1}-a_\pm},\,\, <t|\theta|t'>=\theta(t-t'), \\\nonumber
&& <t|\theta^{-1}|t'>=-\frac{d}{dt}\delta(t-t'),
a_i(t)=iA_{i,\mu}(x(t))\frac{d}{dt}x_\mu(t).
\end{eqnarray}
In the $w[\psi,\psi^\dagger]$ the measure of the integration has a factorized form 
$ \prod_{\pm}^{N_{\pm}}\frac{d\zeta_\pm}{\bar V_{\pm}[\psi^{\dagger} ,\psi ]}$ as in the Eq.~(\ref{PobEq}). It provide the way for the 
extension of this Eq.. Extended Eq. with the account of the light quarks has a form: 
\begin{eqnarray}\label{Eqw}
&&w{-1}[\psi,\psi^\dagger]=\\\nonumber
&&= \theta^{-1} +\int  \prod_{\pm}^{N_{\pm}}\frac{d\zeta_\pm}{\bar V_{\pm}[\psi^{\dagger} ,\psi ]}  \sum_i\left(w[\psi,\psi^\dagger]-\hat A^{-1}_i\right)^{-1}.
\end{eqnarray}
Again, we have the approximate solution of this Eq. as: 
\begin{eqnarray}
&&w^{-1}[\psi,\psi^\dagger]-\theta^{-1}=
\\
&&= \frac{N}{2}\sum_\pm \int\frac{ d\zeta_\pm}{\bar V_{\pm}[\psi^{\dagger} ,\psi ]} V_{\pm}[\psi^{\dagger} ,\psi ]\left( \theta-a_\pm^{-1}\right)^{-1}+ O(N^2/V^2)
\nonumber
\\
&&=-\frac{N}{2}\sum_{\pm}\int\frac{ d\zeta_\pm}{\bar V_{\pm}[\psi^{\dagger} ,\psi ]}V_{\pm}[\psi^{\dagger} ,\psi ]\frac{1}{\theta}
(w_\pm-\theta)\frac{1}{\theta}+ O(N^2/V^2)
\nonumber
\\
\nonumber
&&\equiv - \frac{N}{2}\sum_\pm \frac{1}{\bar V_{\pm}[\psi^{\dagger} ,\psi ]}\Delta_{H,\pm}[\psi^{\dagger},\psi ] + O(N^2/V^2)
\end{eqnarray}
and finally we get
\begin{eqnarray}
\label{SH1}
S_H=\left[\frac{1}{\theta^{-1} - \lambda\sum_\pm\Delta_{H,\pm}[\frac{\delta}{\delta\xi} ,\frac{\delta}{\delta\xi^+}] }
 e^{-\xi^+\left(\hat p + i(m+M(p))\right)^{-1}\xi}\right]_{|_{\xi=\xi^+=0}}.
\end{eqnarray}
If to neglect by overlapping quark loops, then 
\begin{eqnarray}
&&S_H^{-1}\approx\left[\left(\theta^{-1} - \lambda\sum_\pm\Delta_{H,\pm}[\frac{\delta}{\delta\xi} ,\frac{\delta}{\delta\xi^+}] \right)
e^{\xi^+\left(\hat p + i(m+M(p))\right)^{-1}\xi}\right]_{|_{\xi=\xi^+=0}}
\nonumber\\
&&=\theta^{-1} - \frac{N}{2VN_c}\sum_\pm\int d^4z_\pm  {\rm tr}_c\left(\theta^{-1}(w_\pm-\theta)\theta^{-1}\right).
\label{SH3}
\end{eqnarray}
The Eq. (\ref{SH3}) exactly coincide with the similar one from~\cite{Diakonov:1989un}.

Now re-write the Eq. (\ref{SH1}) introducing heavy quark fiels $Q,Q^\dagger$:
\begin{eqnarray}
&& S_H=e^{\left[-{\rm tr}\ln\left(\hat p \,+\, i(m+M(p))\right)\right]}\int D\psi D\psi^{\dagger}  D Q D Q^\dagger \,\,Q \, Q^\dagger\\\nonumber
&&\times\exp\left[\psi^{\dagger}(\hat p +i(m+M(p)))\psi+  Q^\dagger\left(\theta^{-1} - \lambda\sum_\pm\Delta_{H,\pm}[\psi^{\dagger},\psi ] \right)Q\right] 
\\\nonumber
&&\times\exp\left[-{\rm tr}\ln\left(\theta^{-1} - \lambda\sum_\pm\Delta_{H,\pm}[\psi^{\dagger},\psi ]\right)\right],
\end{eqnarray}
where last exponent represent the (negligible) contribution of the heavy quark loops, while 
the second one has the heavy and light quarks interaction action, explicitly represented by 
\begin{eqnarray}
&& - \lambda\sum_\pm Q^\dagger\Delta_{H,\pm}[\psi^{\dagger},\psi ]Q=
\nonumber\\
&&= - i\lambda\sum_\pm\int d^4z_\pm \frac{d^4 k_1}{(2\pi)^4} \frac{d^4 k_2}{(2\pi)^4}  e^{(i(k_2-k_1)z_\pm)} (2\pi\rho )^2 F(k_1 )F(k_2 )
\nonumber \\ \nonumber 
&&\times\left[ \frac{1}{N_c^2}\psi^+(k_1)\frac{1\pm\gamma_5}{2}\psi(k_2)Q^+ {\rm tr}_c\left(\theta^{-1}(w_\pm-\theta)\theta^{-1}\right)Q\right.
\\\nonumber 
&&\left.+\frac{1}{32(N_c^2-1)}\psi^+(k_1) (\gamma_\mu\gamma_\nu \frac{1\pm\gamma_5}{2})\lambda^i \psi(k_2){\rm tr}(\tau^{\mp}_{\mu}\tau^{\pm}_{\nu}\lambda^j)\right.
\\
&&\times\left. Q^+ {\rm tr}_c\left(\theta^{-1}(w_\pm-\theta)\theta^{-1}\lambda^j \right)\lambda^i Q\right].
\end{eqnarray}
We see that the heavy-light quarks interactions terms has a form of the product of the colorless currents of a heavy and light quarks together with similar term of the colorful currents product. The structure of these currents are defined by the instanton color orientation integration,
 while the instanton position integration provide energy-momentum conservation in the interaction vertex. 
 
 At the $N_f>1$ case we have an interaction vertex with $N_f$ pairs of a light quark legs and the pair of a heavy quark legs. The specific structure of the interaction is defined again by instanton color orientation and will be much more reach then at $N_f=1$ case.  We expect that the action generated by the instantons will have reach symmetry properties related to light and heavy quarks sectors both. Namely, it appear the light-heavy quarks interaction terms leading to the specific traces of the light quarks chiral symmetry in light-heavy quarks systems.
 
\section{Heavy quark anti-quark system.}
\label{Heavy quark antiquark}

Now  it is considered the correlator for this system again with the account of ($N_f=1$ case) light quark contribution:
\begin{eqnarray}
&&<T|C(L_1,L_2)|0>=\frac{1}{Z}\int D\psi D\psi^{\dagger} 
\left\{\prod_{\pm}^{N_{\pm}}\bar V_{\pm}[\psi^{\dagger} ,\psi ]\right\}\\ \nonumber
&&\times\exp\int\left(\psi^{\dagger}(\hat p+im )\psi\right)<T|W[\psi,\psi^\dagger]|0>,
\\ \nonumber
&&<T|W[\psi,\psi^\dagger]|0>
=\int\frac{ D\zeta}{\left\{\prod_{\pm}^{N_{\pm}}\bar V_{\pm}[\psi^{\dagger} ,\psi ]\right\}}
\left\{\prod_{\pm}^{N_{\pm}}V_{\pm}[\psi^{\dagger} ,\psi ]\right\}
\\ \nonumber
&&{\rm Tr}<T|\left(\theta^{-1}-\sum_i a^{(1)}_i\right)^{-1}|0>
<0|\left(\theta^{-1}-\sum_i a^{(2)}_i\right)^{-1}|T>.
\end{eqnarray}
Here the correlator is a Wilson loop along the rectangular contour  $L\times r$, where the sides $L_1,L_2$ are parallel to $x_4$ axes and 
separated by the distance $r$. The  $a^{(1)},a^{(2)}$ are the projections of the instantons onto the lines $L_1,L_2.$ 
In the ref.~\cite{Diakonov:1989un} this correlator was considered within the approach similar to the Eq.~(\ref{PobEq}) of the ref.~\cite{Pobylitsa:1989uq} but without a light quarks.
  
  The argumentation, which provided the derivation of the Eq~(\ref{Eqw}), is applicable to the present case and leads to the similar Eq..
 \begin{eqnarray}
&&W^{-1}[\psi,\psi^\dagger]= 
\\\nonumber
&&= w_1^{-1}[\psi,\psi^\dagger]\otimes w_2^{-1,T}[\psi,\psi^\dagger]
-\frac{N}{2}\sum_\pm\int\frac{ d\zeta_\pm}{ \bar V_{\pm}[\psi^{\dagger} ,\psi ]}  
\\ \nonumber
&&\times V_{\pm}[\psi^{\dagger} ,\psi ] \left(w_1[\psi,\psi^\dagger]-a^{(1)-1}_\pm\right)^{-1}\otimes\left(w_2[\psi,\psi^\dagger]-a^{(2)-1}_\pm\right)^{-1,T} , 
\end{eqnarray}  
 where, superscript $T$ means the transposition, $\otimes$ -- tensor product. 
   This Eq. has an approximate solution:  
\begin{eqnarray}
&&W^{-1}[\psi,\psi^\dagger]= w_1^{-1}[\psi,\psi^\dagger]\otimes w_2^{-1,T}[\psi,\psi^\dagger]
\\ \nonumber
&&
-\frac{N}{2}\sum_\pm\int\frac{ d\zeta_\pm}{ \bar V_{\pm}[\psi^{\dagger} ,\psi ]} 
 V_{\pm}[\psi^{\dagger} ,\psi ] 
\\ \nonumber
&&\times\left(\theta^{-1}\left(w^{(1)}_\pm-\theta\right)\theta^{-1}\right)
\otimes\left(\theta^{-1}\left(w^{(2)}_\pm-\theta\right)\theta^{-1}\right)^{T}+ O(N^2/V^2).
\end{eqnarray}
 and 
\begin{eqnarray}
&&w_1^{-1}[\psi,\psi^\dagger]=\theta^{-1}-
\\\nonumber
&&-\frac{N}{2}\sum_{\pm}\frac{ d\zeta_\pm}{ \bar V_{\pm}[\psi^{\dagger} ,\psi ]} V_{\pm}[\psi^{\dagger} ,\psi ]\theta^{-1}(w^{(1)}_\pm-\theta)\theta^{-1}+ O(N^2/V^2)
\\\nonumber
&&=\theta^{-1} - \frac{N}{2}\sum_\pm \frac{1}{\bar V_{\pm}[\psi^{\dagger} ,\psi ]}\Delta^{(1)}_{H,\pm}[\psi^{\dagger},\psi ] + O(N^2/V^2)
\end{eqnarray}
and similar for the $w_2^{-1}[\psi,\psi^\dagger].$

From previous calculations we see that the lowest orders on $\frac{N}{N_cV}$ in $C(L_1,L_2)$ are given by the integration over $\psi,\psi^\dagger$ of the  $W^{-1}[\psi,\psi^\dagger]. $ Here it was neglected by overlapping quark loops. 
Then, we have the new interaction term between heavy quarks located on the lines $L_1$ and $L_2$ due to exchange of 
the light quarks between them.
Explicitly the integration of the first term in $W^{-1}[\psi,\psi^\dagger]$ over $\psi,\psi^\dagger$ leads to:
\begin{eqnarray}
&&\frac{1}{Z}\int D\psi D\psi^{\dagger} 
\left\{\prod_{\pm}^{N_{\pm}}\bar V_{\pm}[\psi^{\dagger} ,\psi ]\right\}\exp\int\psi^{\dagger}(\hat p+im )\psi
\\\nonumber
&&\times\, w_1^{-1}[\psi,\psi^\dagger]\otimes w_2^{-1,T}[\psi,\psi^\dagger]
=\left(\theta^{-1}-\lambda\sum_\pm\Delta^{(1)}_{H,\pm}[\frac{\delta}{\delta\xi} ,\frac{\delta}{\delta\xi^+}]\right)
\\\nonumber
&& \otimes
\left(\theta^{-1}-\lambda\sum_\pm\Delta^{(2)}_{H,\pm}[\frac{\delta}{\delta\xi} ,\frac{\delta}{\delta\xi^+}]\right)^{T} {e^{-\xi^+\left(\hat p \,+\, i(m+M(p))\right)^{-1}\xi}}_{|_{\xi=\xi^+=0}}.
\end{eqnarray}
Light quarks generated potential is given by
\begin{eqnarray}
&&V_{lq}=\left(\lambda\sum_\pm\Delta^{(1)}_{H,\pm}[\frac{\delta}{\delta\xi_1} ,\frac{\delta}{\delta\xi_1^+}]\right) \otimes
\left(\lambda\sum_\pm\Delta^{(2)}_{H,\pm}[\frac{\delta}{\delta\xi_2} ,\frac{\delta}{\delta\xi_2^+}]\right)^{T} 
\nonumber\\
&&\times {e^{\left[-\xi_2^+\left(\hat p \,+\, i(m+M(p))\right)^{-1}\xi_1-\xi_1^+\left(\hat p \,+\, i(m+M(p))\right)^{-1}\xi_2\right]}}|_{\xi=\xi^+=0}.
\end{eqnarray}
The range of this potential is controlled by dynamical light quark mass $M\sim 350$~MeV and might be important for the heavy quarkonium states properties.
\section{ Conclusion.}
\label{ Conclusion}

Approximating the gluon field by the instanton configurations
it was derived the low-frequency part of the light quark determinant in
the presence of quark sources.
It was provided the calculation of the instanon generated light-heavy quarks interaction terms and 
the heavy quark propagator with the account of the light
quark determinant together with the QCD instanton vacuum properties at the $N_f=1$ case. 
 With these knowledge it was calculated the light quark contribution to
the interaction between heavy quarks. 
The extension of this approach to $N_f>1$ case is obvious and provide the possibility for the detailed investigation 
of the role  of the  light quarks chiral symmetry and its spontaneous breaking for the heavy and heavy-light quarks systems. 
The estimations of the light quark contributions to their properties are on the way.


\begin{thebibliography}{69}
 \bibitem{Belle}{[}Belle Collaboration{]}: S.~K.~Choi \emph{et al.},
 Phys.~Rev.~Lett.~\textbf{91} (2003) 262001 ;\\
 K.~Abe \emph{et al.} ,
 Phys.~Rev.~Lett.~\textbf{87} (2001) 091802;\\
 K.~Abe \emph{et al.},
 Phys.~Rev.~D \textbf{66} (2002) 071102.
 
 \bibitem{BABAR}[BABAR Collaboration]:  B.~Aubert \emph{et al.},
 Phys.~Rev.~Lett.~\textbf{87} (2001) 091801;\\ 
 B.~Aubert \emph{et al.},
 Nucl.~Instrum.~Meth.~A \textbf{479} (2002) 1;\\ 
 B.~Aubert \emph{et al.},
 Phys.~Rev.~Lett.~\textbf{90} (2003) 242001. 
 
 \bibitem{K2K}[K2K Collaboration]: M.~H.~Ahn \emph{et al.},
 Phys.~Rev.~Lett.~\textbf{90} (2003) 041801;\\
 M.~Hasegawa \emph{et al.} {[},
 Phys.~Rev.~Lett.~\textbf{95} (2005) 252301 {[};\\
R.~Gran \emph{et al.} {[},
 Phys.~Rev.~D~\textbf{74} (2006) 052002.
 
 \bibitem{MiniBooNE}{[}MiniBooNE Collaboration{]}: 
 A.~A.~Aguilar-Arevalo \emph{et al.} ,  Phys.~Rev.~Lett.~\textbf{100} (2008) 032301;\\
A.~A.~Aguilar-Arevalo \emph{et al.}, Phys.~Lett.~B~\textbf{664} (2008).
 
 \bibitem{NuTeV} {[}NuTeV Collaboration{]}: G.~P.~Zeller \emph{et al.},
 Phys.~Rev.~Lett.~\textbf{88} (2002) 091802 {[}Erratum-ibid.~\textbf{90}
 (2003) 239902{]};\\
 M.~Goncharov \emph{et al.},
 Phys.~Rev.~D \textbf{64} (2001) 112006;\\
 A.~Romosan \emph{et al.},
 Phys.~Rev.~Lett.~\textbf{78} (1997) 2912. 
 
 \bibitem{Minerva}[Minerva Collaboration]: D.~Drakoulakos \emph{et al.}, arXiv:hep-ex/0405002;\\ 
 K.~S.~McFarland {[},
 Nucl.~Phys.~Proc.~Suppl.~\textbf{159} (2006) 107. 
 
 
 \bibitem{Eichten:1979ms}E.~Eichten, K.~Gottfried, T.~Kinoshita,
 K.~D. Lane and T.~M.~Yan, Phys.~Rev.~D \textbf{21} (1980) 203. 
 
 \bibitem{Bodwin:1994jh}G.~T.~Bodwin, E.~Braaten and G.~P.~Lepage,
 Phys.~Rev.~D \textbf{51} (1995) 1125 {[}Erratum-ibid.~D \textbf{55}
 (1997) 5853{]}. 
 
 \bibitem{Isgur:1989}N.~Isgur and M.~B.~Wise, Phys.~Lett.~B
 \textbf{232} (1989) 113;\\ 
 N.~Isgur and M.~B.~Wise, Phys.~Lett.~B
 \textbf{237} (1990) 527. 
 
 \bibitem{Diakonov} D.~Diakonov and V.~Y.~Petrov, 
 Nucl.\ Phys.\ B \textbf{245} (1984) 259;\\
 D.~Diakonov, ~V.~Polyakov and C.~Weiss,
 Nucl.\ Phys.\ B \textbf{461} (1996) 539;\\
D.~Diakonov, Prog.\ Part.\ Nucl.\ Phys.\ \textbf{51},
 173 (2003).
 
 \bibitem{Pobylitsa:1989uq} P.~V.~Pobylitsa, 
  Phys.\ Lett.\ B \textbf{226} (1989) 387. 
 
 \bibitem{Diakonov:1989un} D.~Diakonov, V.~Y.~Petrov and P.~V.~Pobylitsa,
  Phys.\ Lett.\ B \textbf{226} (1989) 372. 
 
 \bibitem{Chernyshev:1995gj} S.~Chernyshev, M.~A.~Nowak and I.~Zahed,
  Phys.\ Rev.\ D \textbf{53} (1996) 5176.
 
 \bibitem{Musakhanov}M.~M.~Musakhanov and F.~C.~Khanna,
 Phys.\ Lett.\ B \textbf{395} (1997) 298;\\
 E.~D.~Salvo and M.~M.~Musakhanov, Eur.\ Phys.\ J.\ C
 \textbf{5}(1998)501;\\ 
  M.~Musakhanov, 
 Eur.\ Phys.\ J.\ C \textbf{9} (1999) 235 ;\\
 M.~Musakhanov, Nucl.\ Phys.\ A \textbf{699}(2002) 340;\\
 M.~M.~Musakhanov and H.~C.~Kim, Phys.\ Lett.\ B
 \textbf{572} (2003) 181;\\
H.~C.~Kim, M.~Musakhanov and M.~Siddikov,
 Phys.\ Lett.\ B \textbf{608} (2005) 95;\\
 H.~C.~Kim, M.~M.~Musakhanov and M.~Siddikov,
  Phys.\ Lett.\ B \textbf{633} (2006) 701;\\
 K.~Goeke, M.~M.~Musakhanov and M.~Siddikov,
  Phys.\ Rev.\ D \textbf{76} (2007) 076007;\\
K.~Goeke, H.~C.~Kim, M.~M.~Musakhanov, M.~Siddikov, 
  Phys.\ Rev.\ D \textbf{76} (2007) 116007;\\
K.~Goeke, M.~Musakhanov, M.~Siddikov,
  Phys.\ Rev.\ D \textbf{81} (2010) 054029.
 
 \bibitem{Leutwyler:2001hn} H.~Leutwyler, 
  Czech.\ J.\ Phys.\ \textbf{52} (2002) B9.
 
 
 \bibitem{Shuryak:1981ff} E.~V.~Shuryak, 
 Nucl.\ Phys.\ B \textbf{203} (1982) 93. 
  
 \bibitem{lattice} M.~C.~Chu, J.~M.~Grandy, S.~Huang and J.~W.~Negele,
 Phys.\ Rev.\ D \textbf{49} (1994) 6039;\\
 J.~W.~Negele, Nucl.\ Phys.\ Proc.\ Suppl.\ \textbf{73}
 (1999) 92;\\
 T.~DeGrand, Phys.\ Rev.\ D \textbf{64}
 (2001) 094508;\\
 P.~Faccioli and T.~A.~DeGrand, Phys.\ Rev.\ Lett.\ \textbf{91}
 (2003) 182001;\\
  P.~O.~Bowman, U.~M.~Heller, D.~B.~Leinweber,
 A.~G.~Williams and J.~b.~Zhang, Nucl.\ Phys.\ Proc.\ Suppl.\ \textbf{128}
 (2004) 23.
 
 \bibitem{Schafer:1996wv} T.~Schafer and E.~V.~Shuryak, Rev.\ Mod.\ Phys.\ \textbf{70}
 (1998) 323.
 
\end{thebibliography}
\end{document}